\documentclass[mathleft
]{an}
\usepackage{graphicx}
\usepackage{times}
\usepackage{txfonts}
\usepackage{natbib}
\begin{document}

\Pagespan{210}{213}
\Yearpublication{2009}%
\Yearsubmission{2008}%
\Month{02}%
\Volume{330}%
\Issue{22}%

\title{A survey of Low Luminosity Compact sources
}

\author{M. Kunert-Bajraszewska\inst{1}\fnmsep\thanks{Corresponding author:
  \email{magda@astro.uni.torun.pl}\newline}
\and  Peter Thomasson \inst{2}
}
\titlerunning{LLC sources}
\authorrunning{Kunert-Bajraszewska \& Thomasson}
\institute{
Toru\'n Centre for Astronomy, N. Copernicus University,
87-100 Toru\'n, Poland
\and 
Jodrell Bank Observatory, The University of Manchester, Macclesfield,
Cheshire SK11 9DL, UK
}

\received{2008 Dec8}
\accepted{2008 Dec18}
\publonline{2009 Feb 15}


\abstract{%
Based on the FIRST and SDSS catalogues
a flux density limited sample of weak Compact Steep
Spectrum (CSS) sources with radio luminosity below $10^{26}\, {\rm W~Hz^{-1}}$ 
at 1.4 GHz has been constructed.
Our previous multifrequency observations of CSS sources have shown
that low luminosity small-scale objects can be strong candidates for compact
faders. This finding supports the idea that some small-size radio
sources are short-lived phenomena because of a lack of significant fuelling.
They never 'grow up' to become FR\,I or FR\,II objects.
This new sample marks the start of a systematical study of the
radio properties and morphologies of the population of low luminosity compact 
(LLC) objects.
An investigation of this new sample should also lead to a better
understanding of compact faders.
In this paper, the results of the first stage of the new project - the
L-band MERLIN observations of 44 low luminosity CSS sources are presented.}

\maketitle

\section{Introduction}

The Gigahertz-Peaked Spectrum (GPS) and Compact Steep Spectrum (CSS)
sources form a well defined classes of radio objects
whereas GPS sources are considered to be entirely contained within the extent of the narrow-line
region ($\leq$ 1~kpc). CSS sources are thought to extend to the size of
the host galaxy ($\leq$ 20~kpc).
GPS/CSS sources are considered to be young radio sources which evolve
into large radio objects during their lifetimes. This interpretation of
the CSS class has now become part of a standard model \citep{fanti90,f95},
and \citet{r96} have proposed an evolutionary scheme
unifying the three classes of sources: GPS, CSS and Large Symmetric Objects
(LSOs).

\citet{sn99} discussed many aspects of the evolutionary
scheme presented above. In particular they concluded that the radio 
luminosities of GPS sources
increase as they evolve, reach a maximum in the CSS phase
and then gra\-dually decrease as these objects grow further to become LSOs.
In fact, LSOs are also divided into two distinct morphological groups of objects: 
FR\,Is and FR\,IIs \citep{fr74}, and there is a relatively sharp luminosity
boundary between them. The nature of the FR-division is still an open
issue, just as are the detailes of the evolutionary process in which CSS sources
become LSOs. It is unclear whether
FR\,II objects evolve to become FR\,Is, or whether a division has
already occurred amongst CSS sources and some of these then become FR\,Is 
and some FR\,IIs. A majority of CSS sources known to date
have high radio luminosities and, if unbeamed, have FR\,II structures. 
It seems reasonable
to suspect that some of the CSSs with lower radio luminosity could be
the progenitors of less luminous FR\,I objects.
According to a recently developed model by \citet{kb07}, all 
sources start out with a FR\,II morphology. Some of them have weaker jets 
which disrupt within the dense environment of their host galaxies and develop
the turbulent lobes of FR\,Is or become hybrid objects
\citep{gopal00,gawron06}.
In the model of \citet{alex00}, sources with disrupted jets fade away. 

The above considerations touch another aspect of the theory of
evolution of radio-loud AGNs. It has already been pointed out by some authors
\citep{sn99,mar03,gug05,kun06}
that there exists a group of GPS/CSS sources that will never reach the LSO
stage, at least in a given cycle of activity if it is recurrent.
Radio galaxies that are no longer fuelled and so
remain in a coasting phase are sometimes termed 'faders'
and so there is no reason why a class of small-scale objects that resemble
large-scale faders should not exist.
Strong support for this idea comes from \citet{rb97}
who proposed a mo\-del in which extragalactic radio sources are intermittent
on timescales of $\sim$$10^4$--$10^5$~years.
This scenario conforms to earlier predictions that Compact Symmetric
Objects (CSOs) could switch off after a short period
of time \citep{r94}. It also solves another well-known problem namely
that the number of compact sources is too high relative
to the general population of radio galaxies \citep{fanti90,odea97}.

Recent observations of weaker compact sources have 
shown that young FR\,Is \citep{gir05} and small-scale fa\-ders
\citep{ kun05,kun06,mar06} do indeed exist. 

\begin{figure}
\centering
\includegraphics[width=6cm,height=8cm]{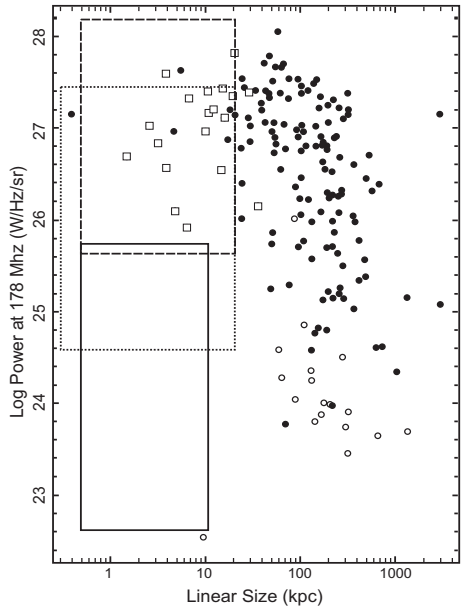}
\caption{Luminosity-size diagram from \citet{aku95} for the revised 3C
sample of \citet{laing83}. Squares indicate CSS sources, filled circles
indicate FR\,II objects, open circles indicate FR\,Is. The
boundaries of the latest samples of CSS sources are also shown: the dashed line indicates
\citet{f2001} sample, the dotted line indicates \citet{kun02} sample and the solid
line indicates the current sample of LLC sources.
}
\label{diagram}
\end{figure}

\section{Previous observations and new sample}

Using an early release of the VLA FIRST survey 
\citep{white97}, \citet{marecki03}  selected a flux density limited, complete
sample of 60~candidate sources fulfilling the basic criteria of
the CSS class. The sample was initially observed with MERLIN C-band, 
after which selected groups of objects were futher observed 
with MERLIN, the EVN and the VLBA
\citep{kun02,mar03,kun05,kun06,kun07,mar06}. 
While a number of sources in the sample were shown to have morphologies similar to
the strong, more 'classic' CSS sour\-ces (two symmetric
edge-brightened lobes with well defined hot\-spots, or core-jet morphology),
the observations also revealed the existence of more exotic objects which were
strong candidates for compact faders. A double source, 0809+404, described
in \citet{kun06} was the best example of such an object.  
The VLBA multifrequency observations showed it to
have a diffuse, amorphous structure, devoid of a dominant core and hotspots.

Most of the GPS/CSS sources known to date are powerful objects with a FR\,II
radio morphology. This is a result of a selection effect, since the main
criterion used so far was flux density and not luminosity.
A next step in studies of GPS and CSS
objects is the observation of compact objects with moderate and low
luminosities. It is assumed
that amongst them there exists a class of objects which are relics of young sources that
has been largely neglected to date.
The goal of this project is to study the properties of LLC
objects and the evolution of the compact object population.

\begin{table}
\centering
\caption{Morphological classification of the sources based on the described
observations. Objects with uncertain
classification are denoted with $"$?$"$. Sources indicated with bold type
are shown in Fig.~\ref{images}.}
\label{tlab}
\begin{tabular}{l l}\hline
Type & Sources\\
\hline
Single & 0754+401, 0801+437, 0835+373,\\
       & 0907+049, 0914+114, 0921+143, \\
       & 1359+525A, 1402+415, 1407+363, \\
       & 1411+553, 1418+053, 1521+324, \\
       & 1532+303, 1717+547\\
Core-jet & 0025+006(?), 0846+017, 0931+033,\\
         & 1009+053, {\bf 1140+058}, 1506+345A,\\
         & 1601+528, 1610+407\\
Double-lobed & 0810+077, 0821+321, 0850+024,\\
             & 0851+024, 0914+504A, {\bf 0942+355},\\
             & 1007+142, 1037+302, 1053+505,\\
             & 1154+435A, 1308+451, 1321+045(?),\\
             & 1542+390, 1543+465, 1558+536A,\\
             & 1624+049, 1715+499\\
Other & {\bf 0854+210}, 0923+079, {\bf 1156+470},\\
      & 1550+444, 1641+320\\
\hline
\end{tabular}
\end{table}

The following criteria were used for the selection of
candidates which were weaker CSS sources.

\begin{itemize}
\item {Using the final release of FIRST, combined with the GB6 4.85\,GHz
survey, unresolved and isolated objects i.e. more compact than the FIRST
beam (5\farcs4) and surrounded by an empty field (field radius 1\arcmin)
were identified.}
\item {The redshifts of the sources, which were
identified with the radio objects, had to be known. These were ex\-trac\-ted from NED and
SDSS/DR4. Consequently, thus providing a low luminosity
criterion: $ L_{1.4{\rm GHz}} < 10^{26}$\,${\rm W~Hz^{-1}}$.}
\item{The flux density was chosen to be in the range 
70\,mJy $\leq S_{1.4{\rm GHz}} \leq$ 1\,Jy in order to produce a sample of manageable 
size, but also to exclude
objects with flux densities too low to be detected in a snapshot
observations.}
\item {We established the steepness criterion: $\alpha_{1.4 {\rm GHz}}^{4.85 {\rm GHz}} >0.7$
($S\propto\nu^{-\alpha}$).}
\end{itemize}

\begin{figure*}
\centering
\includegraphics[width=7cm,height=7cm]{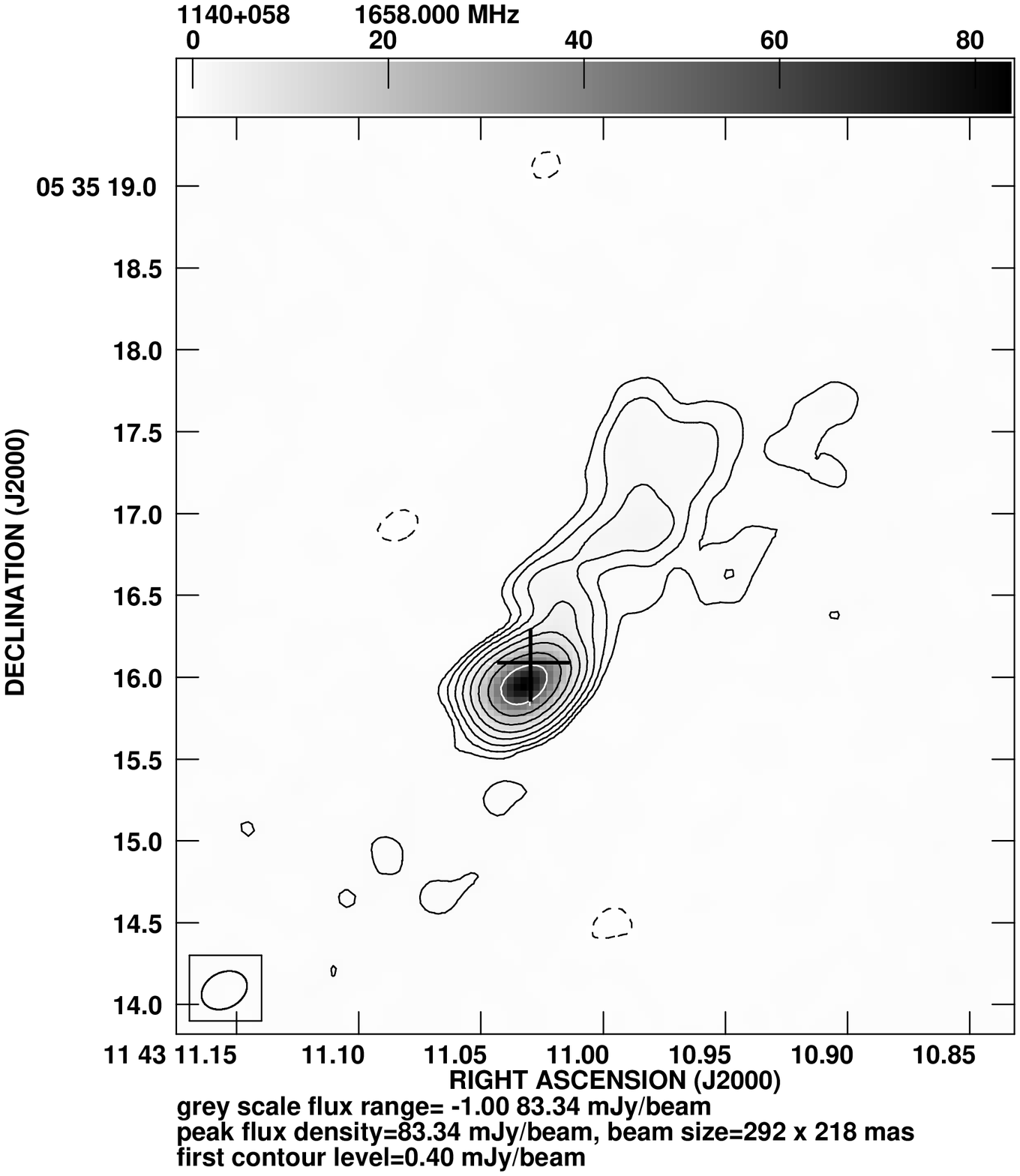}
\includegraphics[width=7cm,height=7cm]{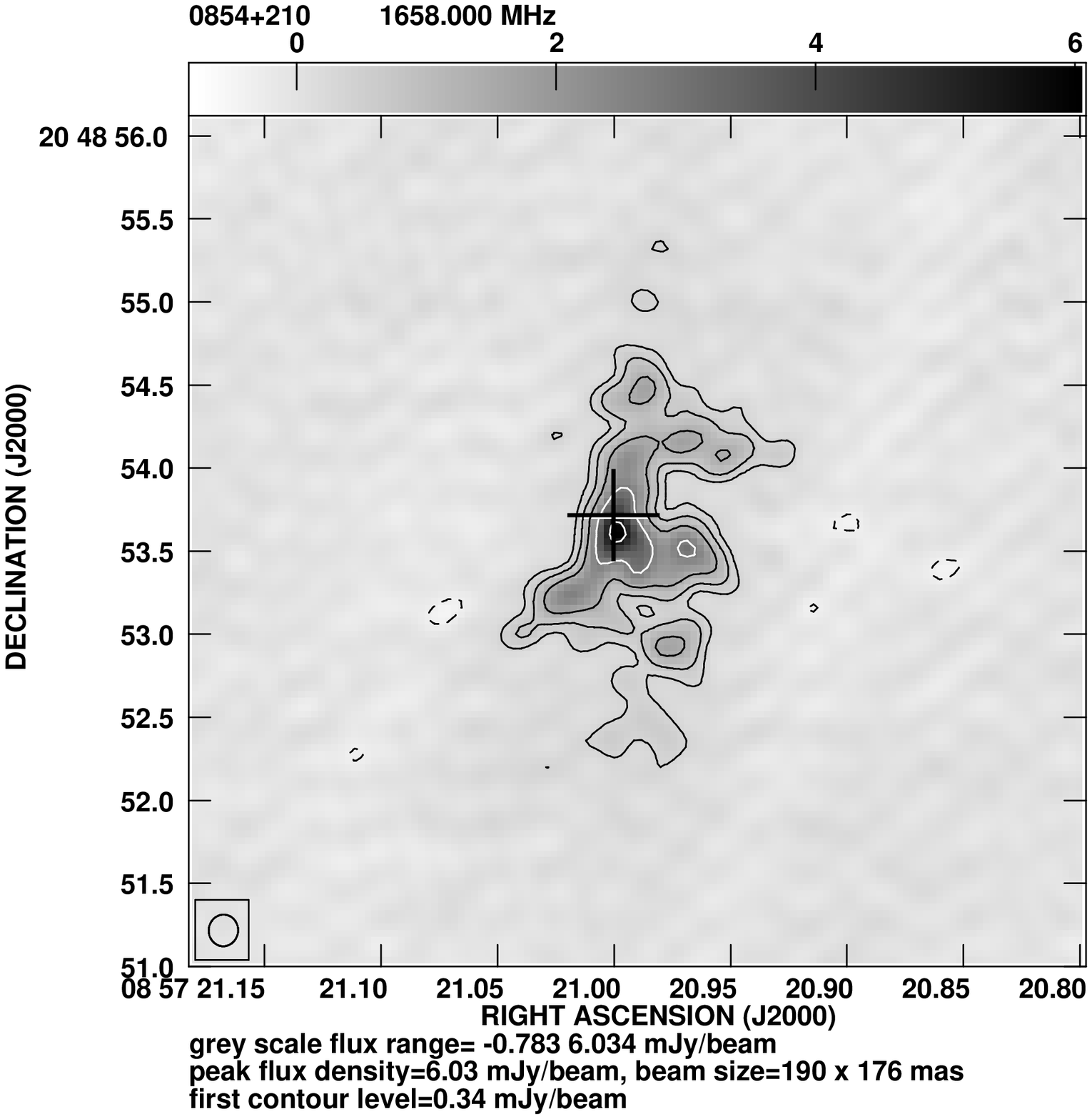}
\includegraphics[width=7cm,height=7cm]{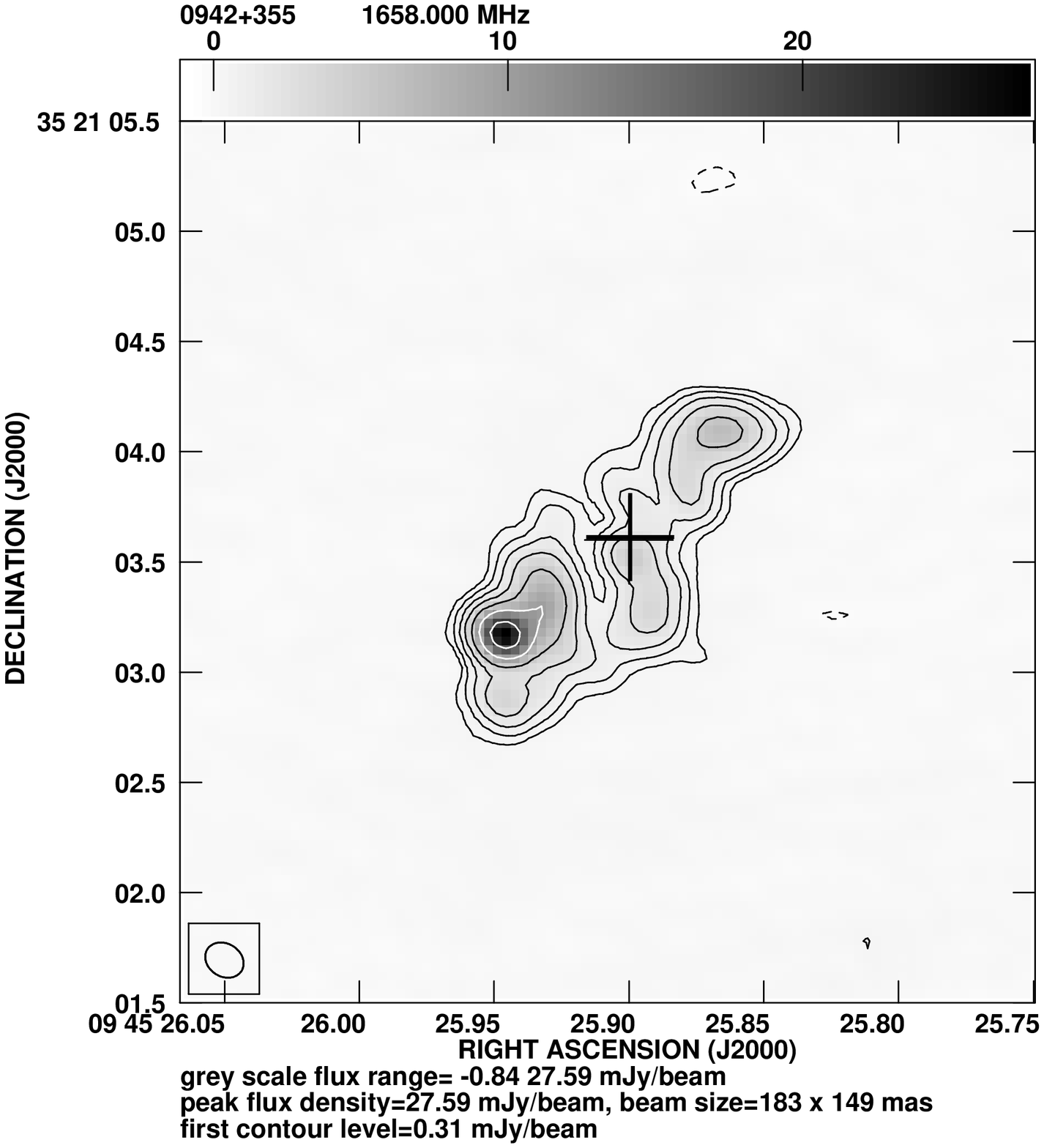}
\includegraphics[width=7cm,height=7cm]{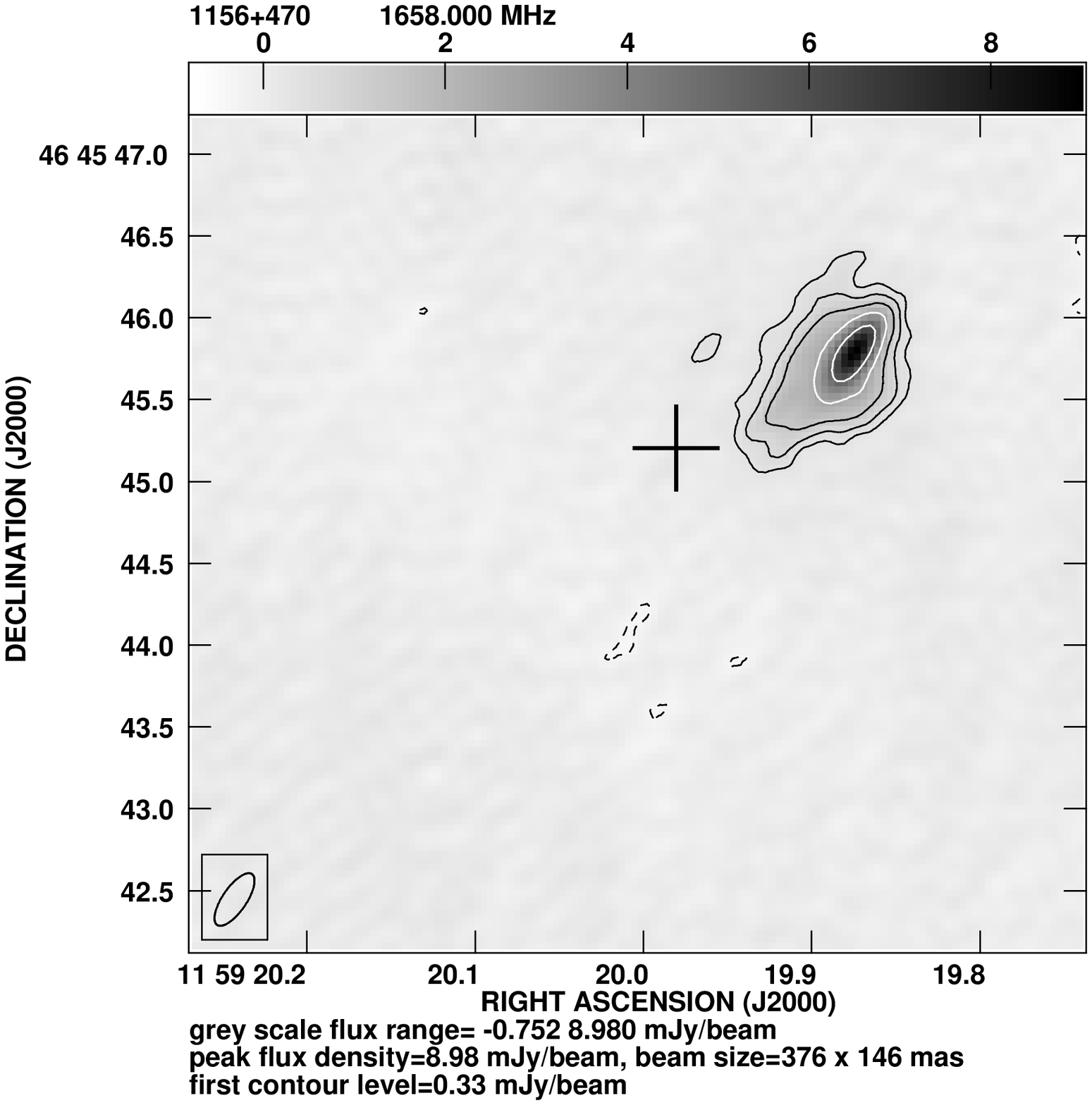}
\caption{MERLIN L-band images of four sources from our new sample.
The selected objects are indicated with bold type in Table~1. Contours increase
by a factor 2, and the first contour level corresponds to $\approx 3\sigma$.
A cross indicates the position of an optical object found using the
SDSS/DR6.
}
\label{images}
\end{figure*}

There is no overlap between the new sample and our
previous one \citep{kun02}.
Finally, the new sample consists of 44 sources.

The initial survey was undertaken using MERLIN at 1.6\,GHz. 
Snapshot observations of the sample were made on
seven separate days between December 2006 and May 2007. 
Each target source together with its associated phase reference 
sources was
observed for $\sim$60\,min including telescope drive times. 
The phase calibrator sources 
were observed twice per
target scan for 1-2 min. Initial amplitude calibration
was derived from daily observations of the unresolved source
OQ208. The preliminary data reduction 
was made using an AIPS-based PIPELINE procedure developed
at JBO. The resulting phase-calibrated images created with PI\-PE\-LI\-NE
were further improved using several cycles of self-ca\-li\-bra\-tion and
imaging, the final maps being produced using IMAGR.\\
Throughout the paper, we assume a cosmology with
${\rm H_0}$=\\
100${\rm\,km\,s^{-1}\,Mpc^{-1}}$, $q_{0}$=0.5.

\section{Preliminary results}

The selection criteria used for the new sample of candidate CSS
sources means that objects with low luminosities that have previously 
never been observed so widely, have been selected.
This is shown on the luminosity-size diagram (Fig.~\ref{diagram}),
originally derived by \citet{aku95} for the revised 3C sample by
\citet{laing83}. 
Now indicated are the samples of \citet{f2001}, \citet{kun02} and the present
sample.
The 178\,MHz flux densities of sources have been estimated for the three
samples using the known spectral indices between 0.4-1.4\,GHz
\citep{f2001} and 1.4-4.9\,GHz (\citet{kun02}, new sample) and assuming
that the indices remain valid down to 178\,MHz. This assumption may not
always be true, but it is sufficient to show the boundaries of the
above-mentioned samples. Approximately one third of the CSS sources from the new
sample have a value for the 178\,MHz luminosity lower than the luminosity
bo\-un\-da\-ry found for FR sources \citep{fr74}, which means that they are compact young
sources with luminosities comparable to FR\,Is.  

About 70$\%$ of the sources from the sample are galaxies and all of them are
nearby objects with redshifts, {\it z}, less than 1, and linear
sizes less than 15\,kpc; i.e. typical for CSS objects. 
Based upon preliminary radio images at 1.6\,GHz, the new sources have been
divided into four categories (Table 1). 'Single' means the source is a pointlike
object unresolved with MERLIN at the observed frequency. 'Core-jet' is a
source with a bright central component and a one- or two-sided jet.
'Double-lobed' objects show at least two kinds of morphologies. Some of them
are compact doubles without a visible radio core at 1.6\,GHz, whereas the
others show
a central weak component and distorted extended structure. Most of the
double objects also show brightness asymmetry. The last category, 'other',
comprises five sour\-ces with peculiar morphologies: two objects are binary
systems with distorted radio morphologies, the other three have only a
single visible lobe.
Fig.~\ref{images} shows example sources from each category except those
which are unresolved.

The large precentage of double objects with an asymmetric morphology in the new
sample confirms a previous finding for the CSS source population.
\citet{saikia01} have shown that,
as a class, the CSSs are more asymmetric than larger radio
sources of similar power. There are at least two explanations of this
observational fact.
There is a high probability that CSS sources  
interact with an asymmetric me\-dium in the central regions of the 
host elliptical galaxy. 

Another explanation is the light-travel time effect, i.e. the time lag between
the images of the lobes as the observer perceives them \citep{ocp98}. 
It could be that the far side of a source is being viewed at an earlier time
than the near side and that for some double asymmetric
sources, the near side hotspot can be seen at a time when it is
being supplied by jet material, whereas the far side kpc-scale
hotspot isn't. Hotspots not being supplied by jet material fade very fast
because electrons diffuse to regions of lower density and
magnetic field and consequently radiate inefficiently.
As a result, such a source may be detected as a single-lobed object, similar
to the three already found in the new sample (Fig.~\ref{images}).
This kind of objects have been already
observed in our previous sample \citep{mar03}, and amongst large-scale sources
\citep{sub03}.

According to earlier \citep{alex00} and recent \citep{kb07} 
theoretical predictions the distorted structure of a source arises when it
tries to escape from the core radius of its host galaxy. Those
sources that survive their evolution through the densest inner
regions of their galactic atmosphere should appear on the smallest
scales as classical double (FR\,II type) objects.
Sources that are disrupted are likely to
have a different morphology. In the model of \citet{kb07} they become
FR\,I sources or hybrid objects. However, \citet{alex00} shows that sources with
disrupted jets are not expected to evolve into FR\,I objects, but fade
so strongly that they become virtually undetectable. 
It is possible that some of the sources from this new sample, namely
'single-lobed' objects, are in a coasting phase now.

\section{Conclusions and future plans}

We have presented some of the results from MERLIN L-band observations of
a new sample of low luminosity compact sources, the details of which
will be discussed in a separate paper.

We have also obtained new polarisation data and MERLIN C-band
observations of some of the sources, which should help to explain the 
asymmetry of the radio structure seen in some of them. Since most of the 
sources have known spectra, it is hoped that further analysis will lead to 
evidence for a lack of activity or interaction with the surrounding
medium. The results of this will also be presented in a forthcoming paper.

\acknowledgements
\item{MERLIN is a UK National Facility operated by
the University of Manchester on behalf of STFC.}
\item{We thank A.Siemiginowska for reading of the paper and a number of
suggestions.}
\item{This work has benefited from research funding from the European
Community's sixth Framework Programme under RadioNet R113CT 2003 5058187.}


\end{document}